\documentclass[prb,twocolumn,showpacs,superscriptaddress,groupedaddress]{revtex4-1}
\usepackage{amsmath}
\usepackage{epsfig,color}
\usepackage{dcolumn}
\usepackage{amsfonts}
\usepackage{amssymb}
\usepackage{graphicx}

\begin{document}

\title{Two-band model of Raman scattering on iron pnictide superconductors}
\author{C. S. Liu$^{1,2}$ and W. C. Wu$^1$}

\affiliation{$^1$Department of Physics, National Taiwan Normal University, Taipei 11677,
Taiwan\\
$^2$Department of Physics, Yanshan University, Qinhuangdao 066004, China}

%\author{C. S. Liu}
%\affiliation{Department of Physics, National Taiwan Normal University, Taipei 11677, Taiwan}
%\affiliation{Department of Physics, Yanshan University, Qinhuangdao 066004, China}
%\author{W. C. Wu}
%\affiliation{Department of Physics, National Taiwan Normal University, Taipei 11677, Taiwan}

\date{\today}

\begin{abstract}
Based on a two-band model, we study the electronic Raman scattering intensity
in both normal and superconducting states of iron-pnictide superconductors.
For the normal state, due to the match or mismatch of the symmetries between band hybridization
and Raman vertex, it is predicted that overall $B_{1g}$ Raman intensity should be much
weaker than that of the $B_{2g}$ channel. Moreover, in the non-resonant regime,
there should exhibit a interband excitation peak at frequency
$\omega\simeq 7.3 t_1 (6.8t_1)$ in the $B_{1g}$ ($B_{2g}$) channel.
For the superconducting state,
it is shown that $\beta$-band contributes most
to the $B_{2g}$ Raman intensity as a result of multiple effects of
Raman vertex, gap symmetry, and Fermi surface topology.
Both extended $s$- and $d_{xy}$-wave pairings in
the unfolded BZ can give a good description to the reported $B_{2g}$ Raman data
[Muschler {\em et al.}, Phys. Rev. B. {\bf 80}, 180510 (2009).], while
$d_{x^2-y^2}$-wave pairing in the unfolded BZ seems to be ruled out.
%We have also give the predicted $B_{1g}$ Raman spectra for testing.
\end{abstract}

\maketitle

\section{Introduction}

Recently superconductivity has been observed in several classes of
iron-pnictide materials. \cite
{kamihara, Wang08, Hsu} Despite that different classes of pnictides have
somewhat different crystal structures, their electronic structures behave
to be quite similar, as confirmed by ARPES and magneto-oscillation measurements.
\cite{Liu177005,Evtushinsky,Coldea}
%Fermi surface consists of two small hole pockets centered around the $
%\Gamma=(0,0)$ point and two small electron pockets centered around the $
%M=(\pi,\pi)$ point.
Band structure calculations indicate that iron pnictides exhibit a
quasi-two-dimensional electronic structure. Their parent compounds are
metals which display an antiferromagnetic long-range order. Superconductivity
is induced either by hole doping or electron doping when part of $\mathrm{Fe}^{2+}$
ions are replaced by $\mathrm{Fe}^{+}$, or
solely by the application of high pressure.
Overall speaking, iron-pnictide shows
a good resemblance with the high-$T_c$ cuprate superconductor,
and hence is an ideal candidate for studying the superconducting (SC)
mechanism of high $T_c$.

One crucial issue towards identifying the interaction that drives superconductivity
is the symmetry structure of Cooper pairs.
A conclusive observation of the pairing symmetry remains unsettled for iron pnictides,
however, Both nodal and nodeless order parameters were reported in
experimental observations. ARPES measurements
clearly indicated a nodeless gap at all points of the Fermi surface.
\cite{Ding_EPL,zhao} Moreover, magnetic penetration depth measurement has
revealed a $T^2$ behavior down to
$0.02T_{c}$ -- a possible signature for the unconventional
$s_{\pm}$ state.\cite{mazin:057003}
The $s_{\pm}$ state is currently a promising pairing candidate for iron pnictides,
which has a sign reversal between $\alpha$ and $\beta$ bands
and can be naturally explained by the spin fluctuation mechanism.
\cite{mazin:057003, wang:047005, Tsuei2010}
On the other hand the neutron-scattering spectrum seems to
suggest the order parameter being fully gapped but without a sign reversal
(called $s_{++}$-wave in contrast to $s_{\pm}$-wave).
\cite{PhysRevB.81.060504, PhysRevB.81.180505}
On the contrary, both scanning SQUID microscopy\cite{hicks-2008} and
NMR measurements \cite{Matano,Grafe08} are
in favor of nodal SC order parameters.
For phase sensitive experiments however, one
point-contact spectroscopy reported was in favor of a nodal gap
\cite{Shan}, while the others reported were in favor of a
nodeless gap. \cite{chen}

Compared to other spectroscopy experiments,
Raman scattering has a unique feature, namely, by manipulating
the polarization directions of incident and scattered photons with
respect to the crystallographic directions,
it can selectively excite quasiparticles (QPs)
on different parts of the Brillouin zone (BZ) . Therefore Raman scattering is
considered a very useful probe for studying pairing symmetry of the SC order
parameter. More explicitly, among various symmetry channels
($A_{1g}$, $A_{2g}$, $B_{1g}$, and $B_{2g}$ within a $D_{\rm 4h}$ symmetry group),
powers laws of low-energy Raman
intensity as well as peak positions can often give
useful information on the pairing symmetry of a superconductor.
Raman scattering had been very
successful on the studies of the high-$T_c$ cuprate superconductors. \cite{RevModPhys.79.175}
It is also anticipated that Raman scattering will give very useful
signals on the iron-pnictide superconductors.

Electronic Raman scattering experiment
has recently been carried out
on high-quality single-crystal
Ba(Fe$_{1-x}$Co$_x$)$_2$As$_2$. \cite{PhysRevB.80.180510}
Both normal- and superconducting-state data were taken and for frequency
below 300 cm$^{-1}$,
it showed that a significant change of the Raman response between
the normal and superconducting state occurs only in
the $B_{2g}$ channel to which a strong peak develops at $\omega=69$cm$^{-1}$ in the
superconducting state. Moreover,
the low-energy spectral weight in the SC state suggests
a gap having nodes for certain doping levels of Ba(Fe$_{1-x}$Co$_x$)$_2$As$_2$.
%Nesting properties
%between the $\alpha$ and $\beta$ sheets rather than electron-phonon
%coupling would provide a strongly enhanced dynamic interaction
%between electrons from nearest-neighbor Fe orbital which is indeed close related to
%the cuprates.
It is of importance to perform a more quantitative theoretical fitting on
the Raman data reported in Ref.~[\onlinecite{PhysRevB.80.180510}] and
this is indeed the goal of the present paper.

On the theoretical side, Boyd {\em et al.} \cite{PhysRevB.79.174521} have
calculated the Raman response for iron-pnictides taking into account
multiple gaps on different Fermi sheets. In their calculations, the
Raman vertices are obtained by the expansion of
harmonic functions for a cylindrical FS and the screening effect
due to the long-range Coulomb interaction is considered but leaving out vertex corrections
capturing effects of final state interactions.
Their results give a criterion to distinguish the momentum and frequency dependence of
the SC order parameters by Raman scattering.
In another study, Chubukov and coworkers \cite{PhysRevB.79.220501, Physics.2.46}
have also calculated the Raman response for iron-pnictide superconductors.
By analyzing the vertex corrections for the extended $s$-wave gap symmetry, they
have revealed a collective mode below $2\Delta_0$ in the $A_{1g}$ channel
that may be crucial to unravel the pairing symmetry.
It provides an alternative way to distinguish between various suggested gap
symmetries of the iron-pnictide superconductors.

The electronic structure of iron-pnictide materials is considered to
be more complex than that of the high-T$_c$ cuprates.
The band structure calculations showed that superconductivity
is associated with the Fe-pnictide layer, and the FS
consists of two hole pockets and two electron pockets. \cite{singh:237003,
PhysRevB.77.220506}
The maximum contribution to the density of states near the FS is due to
the Fe-3$d$ orbitals.
Several tight-binding models are thus proposed to construct
the band structure and several competing orders are suggested to exist
in these materials.
\cite{PhysRevB.77.220506, PhysRevLett.101.087004, PhysRevLett.101.057008, 0953-8984-20-42-425203}
In this paper, we shall study the Raman response based on a two-band model.\cite{raghu_prb_08}
This minimal model is considered to be a promising one which captures
the major features of the electronic structures of several classes of
iron-pnictide materials. As mentioned before,
our aim is to give a more quantitative fitting to the current available
Raman data and uncover useful fitting parameters.
It is hoped that our study,
together with other theoretical Raman
works (Refs.~\onlinecite{PhysRevB.79.220501,PhysRevB.79.174521})
can give a more complete picture on the pairing symmetry and electronic structure
of iron pnictides.

%We then calculate the ultrasonic attenuation with the minimal
%two-band model, and find it is sensitive to angle dependance of QP excitation density which is
%relate to band structure directly. So the
%ultrasonic attenuation measurement is an alternative method to detect
%superconductivity of Fe-pnictides.

The paper is organized as the following. In Sec.~\ref{Model Hamiltonian},
we present the two-band model Hamiltonian for iron-pnictide
superconductors. In Sec.~\ref{Raman Scattering},
theoretical derivations are given for the Raman scattering both in the normal
and in the SC states
with respect to the two-band model given in Sec.~\ref{Model Hamiltonian}.
Calculations of Raman response are presented for both normal and SC states.
Some predictions are made.
Especially fitting to the $B_{2g}$ Raman peak in the SC state are performed
with useful fitting parameters given.
Sec.~\ref{Summary} is a summary of this paper.

\section{Model Hamiltonian} \label{Model Hamiltonian}

For iron-pnictides, we start from the minimal two-band model introduced in
Ref.~[\onlinecite{raghu_prb_08}]. The normal-state Hamiltonian reads as
\begin{equation}
H_{0}=\sum_{\mathbf{k}\sigma }\left(
\begin{array}{cc}
c_{\mathbf{k}\sigma }^{\dag } & d_{\mathbf{k}\sigma }^{\dag }%
\end{array}%
\right) \left(
\begin{array}{cc}
\varepsilon _{x}-\mu & \varepsilon _{xy} \\
\varepsilon _{xy} & \varepsilon _{y}-\mu%
\end{array}%
\right) \left(
\begin{array}{c}
c_{\mathbf{k}\sigma } \\
d_{\mathbf{k}\sigma }%
\end{array}%
\right),  \label{normal state Hamiltonian}
\end{equation}%
where $\mu$ is the chemical potential and
$(c_{\mathbf{k}\sigma}^{(\dagger)},d_{\mathbf{k}\sigma}^{(\dagger)})$
are annihilation (creation) operators that annihilate (create) an electron in
the $d_{xz}$- and $d_{yz}$-orbital with spin $\sigma $ and wavevector $
\mathbf{k}$, respectively. Energy dispersions of orbital $d_{xz}$ and $d_{yz}$
are given by $\varepsilon_{x}=-2t_{1}\cos k_x-2t_{2}\cos k_y-4t_{3}\cos k_x \cos k_y$
and $\varepsilon_{y}=-2t_{2}\cos k_x-2t_{1}\cos k_y-4t_{3}\cos k_x \cos k_y$ respectively,
which are coupled by the $d_{xy}$ orbital with dispersion
$\varepsilon _{xy}=-4t_{4}\sin k_x\sin k_y$.
By the unitary transformation
\begin{equation}
\left(
\begin{array}{c}
c_{\mathbf{k}\sigma } \\
d_{\mathbf{k}\sigma }%
\end{array}%
\right) =\left(
\begin{array}{cc}
\cos \theta _{\mathbf{k}} & -\sin \theta _{\mathbf{k}} \\
\sin \theta _{\mathbf{k}} & \cos \theta _{\mathbf{k}}%
\end{array}%
\right) \left(
\begin{array}{c}
\alpha _{\mathbf{k}\sigma } \\
\beta _{\mathbf{k}\sigma }%
\end{array}%
\right) ,  \label{unitary transformation}
\end{equation}%
Hamiltonian (\ref{normal state Hamiltonian}) can be transformed to
\begin{equation}
H_{0}=\sum_{\mathbf{k}\sigma }\left(
\begin{array}{cc}
\alpha _{\mathbf{k}\sigma }^{\dag } & \beta _{\mathbf{k}\sigma }^{\dag }%
\end{array}%
\right) \left(
\begin{array}{cc}
E_{-} & 0 \\
0 & E_{+}%
\end{array}%
\right) \left(
\begin{array}{c}
\alpha _{\mathbf{k}\sigma } \\
\beta _{\mathbf{k}\sigma }%
\end{array}%
\right) ,  \label{diagnalization Hamiltonian}
\end{equation}%
where $E_{\pm}=\varepsilon _{+}\pm \sqrt{\varepsilon _{-}^{2}+\varepsilon
_{xy}^{2}}-\mu $ with  $\varepsilon _{\pm }\equiv\left( \varepsilon _{x}\pm
\varepsilon _{y}\right) /2$.
The new basis $(\alpha _{\mathbf{k}\sigma },\beta _{\mathbf{k}\sigma})$
consists of new fermionic QP operators in the $E_\pm$ bands which are hybrids
of the $d_{xz}$- and $d_{yz}$-orbital. In the literature, $E_-$ ($E_+$) is usually
labeled as the $\alpha$ ($\beta$) band.

% and $\tan 2\theta _{\mathbf{k}}=\varepsilon_{xy}/\varepsilon _{_{\mathbf{-}}}$.
The coherence factors in (\ref{unitary transformation}) can be solved to be
\begin{eqnarray}
\cos ^{2}\theta _{\mathbf{k}} &=&\frac{1}{2}\left( 1+\frac{\varepsilon _{%
\mathbf{-}}}{\sqrt{\varepsilon _{xy}^{2}+\varepsilon _{\mathbf{-}}^{2}}}%
\right) \notag\\
\sin ^{2}\theta _{\mathbf{k}} &=&\frac{1}{2}\left( 1-\frac{\varepsilon _{%
\mathbf{-}}}{\sqrt{\varepsilon _{xy}^{2}+\varepsilon _{\mathbf{-}}^{2}}}%
\right).
\end{eqnarray}%
Moreover, the single-particle Matsubara Green's functions in the normal-state
are
\begin{eqnarray}
\mathcal{G}^0_{cc}(\mathbf{k},ip_{n})&=&-\int_0^\beta d\tau~ e^{ip_n\tau}\langle
c_{{\bf k}\sigma}(\tau)c^\dagger_{{\bf k}\sigma}(0)\rangle_0\nonumber\\
 &=&\frac{\cos ^{2}\theta _{\mathbf{k}}}{%
ip_{n}-E_{-}}+\frac{\sin ^{2}\theta _{\mathbf{k}}}{ip_{n}-E_{+}}, \notag\\
\mathcal{G}^0_{dd}(\mathbf{k},ip_{n}) &=&-\int_0^\beta d\tau~ e^{ip_n\tau}\langle
d_{{\bf k}\sigma}(\tau)d^\dagger_{{\bf k}\sigma}(0)\rangle_0\nonumber\\
&=&\frac{\sin ^{2}\theta _{\mathbf{k}}}{%
ip_{n}-E_{-}}+\frac{\cos ^{2}\theta _{\mathbf{k}}}{ip_{n}-E_{+}}, \\
\mathcal{G}^0_{cd}(\mathbf{k},ip_{n})&=&\mathcal{G}^0_{dc}(\mathbf{k},ip_{n})
=-\int_0^\beta d\tau~ e^{ip_n\tau}\langle
c_{{\bf k}\sigma}(\tau)d^\dagger_{{\bf k}\sigma}(0)\rangle_0\nonumber\\
&=&-\frac{\sin 2\theta _{\mathbf{k}}}{2}\left( \frac{1}{ip_{n}-E_{-}}-\frac{1%
}{ip_{n}-E_{+}}\right).  \notag
\end{eqnarray}%

Throughout this paper, we shall choose $t_{1}=-1$, $t_{2}=1.3$, $%
t_{3}=t_{4}=-0.85$, and $\mu =1.54$, all measured in units of $|t_{1}|$.
These represent a good fit to the first-principle calculations
for the band structures of LaOFeAs. \cite{Xu67002}

\section{Raman Scattering}  \label{Raman Scattering}

\subsection{Normal state}  \label{sec31}
Raman scattering intensity is proportional to the imaginary part of the
effective density-density correlation function $\chi (\mathbf{q},\tau
)=\langle T_{\tau }[\tilde{\rho}(\mathbf{q},\tau ),\tilde{\rho}(-\mathbf{q}%
,0)]\rangle $ in the $\mathbf{q}\rightarrow 0$ limit. For the current
two-band system, effective density operator associated with Raman scattering
can be given by
\begin{eqnarray}
\tilde{\rho}(\mathbf{q},\tau )& \equiv& {\sum_{\mathbf{k}\sigma }}
[\gamma _{\mathbf{k}}^{c}c_{\mathbf{k}+\mathbf{q}\sigma }^{\dag }(\tau )c_{%
\mathbf{k}\sigma }(\tau ) \nonumber\\
& +&\gamma _{\mathbf{k}}^{d}d_{\mathbf{k}+\mathbf{q}\sigma }^{\dag }(\tau
)d_{\mathbf{k}\sigma }(\tau )],
\label{rho_tilde}
\end{eqnarray}%
where $\gamma _{\mathbf{k}}^{c}$ ($\gamma_{\mathbf{k}}^{d}$) is the Raman vertex
associated with the electrons on $d_{xz}$ ($d_{yz}$) orbital.
In the Matsubara frequency space, symmetry-channel dependent
irreducible Raman response function in the normal state has been solved to be
\begin{align}
& \chi_N({\bf q}\rightarrow 0,i\omega_n)={1\over 4}\sum_{\mathbf{k}}\Gamma _{\mathbf{k}}^{2}
\label{Raman response in normal state} \\
& \times \left[ \frac{f(E_-)-f(E_+)}{i\omega
_{n}-(E_+ - E_-)}+\frac{f(E_+)-f(E_-)}{i\omega _{n}-(E_- - E_+)}\right],   \notag
\end{align}%
where $f(E)=1/[\exp(E/{k_B T})+1]$ is the Fermi distribution function and
\begin{equation}
\Gamma _{\mathbf{k}}=\sin 2\theta _{\mathbf{k}}\left( \gamma _{\mathbf{k}%
}^{c}+\gamma _{\mathbf{k}}^{d}\right)
\label{effect vertex of normal state}
\end{equation}
is a weighting factor for the symmetry-dependent
Raman scattering of a two-band system. $\Gamma_{\mathbf{k}}$ is comprised
of two parts where $\sin 2\theta _{\mathbf{k}}$ corresponds to the
effect of orbital hybridization, while $( \gamma _{\mathbf{k}%
}^{c}+\gamma _{\mathbf{k}}^{d})$ is the sum of the Raman vertexes associated
with two individual orbitals. In the case of zero orbital hybridization
($\varepsilon_{xy}\rightarrow 0$),
$\sin 2\theta _{\mathbf{k}}\rightarrow 0$ and hence $\chi_N\rightarrow 0$.

%We ignore the vertex correction due
%to the long-range Coulomb interaction which is particularly important
%in the $A_{1g}$ channel which is not investigated in the current context.

\begin{figure}[ptb]
\begin{center}
\includegraphics[
width=9cm ]{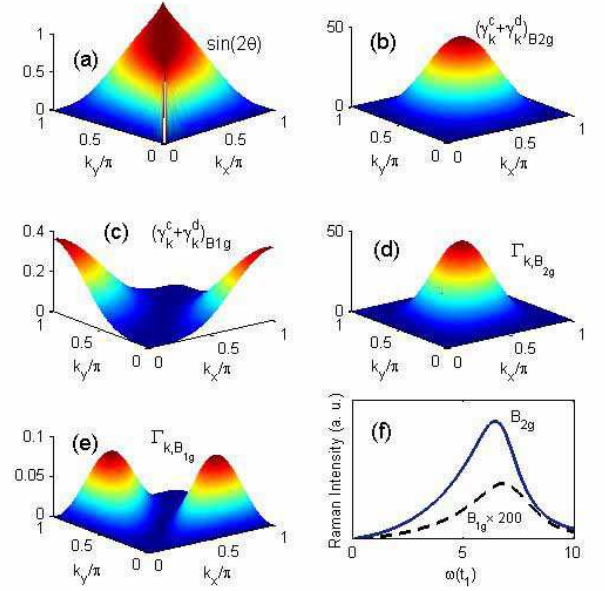}\vspace{-0.5cm}
\end{center}
\caption{(Color online) Panel (a)--(e): To see the properties of the
channel-dependent weighting factor $\Gamma _{\mathbf{k}}$,
$\sin 2\theta _{\mathbf{k}}$, $(\gamma_{\mathbf{k}}^{c}+\gamma _{\mathbf{k}}^{d})_{B_{2g}}$,
$(\gamma_{\mathbf{k}}^{c}+\gamma _{\mathbf{k}}^{d})_{B_{1g}}$,
$\Gamma _{\mathbf{k},B_{2g}}$, and $\Gamma _{\mathbf{k},B_{1g}}$ are plotted respectively
in the first quadrant of the BZ.
Panel (f) is the calculated normal-state Raman intensity.} \label{fig1}
\end{figure}

When the energy of incident light is much smaller than the optical band gap
of the system, the
contribution from the resonant channel is negligible. Consequently, Raman vertex can
be obtained in terms of the curvature of the band dispersion, known as the
inverse effective mass approximation.\cite{PhysRevB.51.16336} That is,
depending on the symmetry of the Raman modes,
$(\gamma _{\mathbf{k}}^{c})_{ij}\sim\partial^2\varepsilon_x/\partial {\bf k}_i\partial {\bf k}_f$
and $(\gamma _{\mathbf{k}}^{d})_{ij}\sim\partial^2\varepsilon_y/\partial {\bf k}_i\partial {\bf k}_f$
with ${\bf k}_i$ and ${\bf k}_f$ being the wavevectors of incident and scattered lights respectively.
In the current two-band model of iron-pnictides, it is obtained that
\begin{eqnarray}
(\gamma _\mathbf{k}^c)_{B_{1g}} &\sim&\frac{1}{2}(\varepsilon_{x}^{xx}
-\varepsilon_{x}^{yy})  =t_{1}\cos{k}_{x}-t_{2}\cos{k}_{y},\nonumber\\
(\gamma _\mathbf{k}^d)_{B_{1g}} &\sim&\frac{1}{2}(\varepsilon_{y}^{xx}
-\varepsilon_{y}^{yy})=t_{2}\cos{k}_{x}-t_{1}\cos{k}_{y},\label{vertex}\\
(\gamma _\mathbf{k}^c)_{B_{2g}}&=&(\gamma _\mathbf{k}^d)_{B_{2g}}\sim\varepsilon_{x}^{xy}
=\varepsilon_{y}^{xy}=4t_{3}\sin{k}_{x}\sin{k}_{y},\nonumber
\end{eqnarray}
where $\varepsilon_{x}^{xx}\equiv\partial^2\varepsilon_x/\partial {\bf k}_x\partial {\bf k}_x$, etc.

In Fig.~\ref{fig1}, we first examine the behaviors of the weighting factor
$\Gamma _{\mathbf{k}}$. Respectively in Fig.~\ref{fig1}(a)--(e), we plot
$\sin 2\theta _{\mathbf{k}}$, $(\gamma_{\mathbf{k}}^{c}+\gamma _{\mathbf{k}}^{d})_{B_{2g}}$,
$(\gamma_{\mathbf{k}}^{c}+\gamma _{\mathbf{k}}^{d})_{B_{1g}}$,
$\Gamma _{\mathbf{k},B_{2g}}$, and $\Gamma _{\mathbf{k},B_{1g}}$
in the first quadrant of the BZ. As shown, both $\sin 2\theta _{\mathbf{k}}$ and
$(\gamma_{\mathbf{k}}^{c}+\gamma _{\mathbf{k}}^{d})_{B_{2g}}$ are peaked at
$(\pi/2,\pi/2)$, while $(\gamma _{\mathbf{k}}^{c}+\gamma
_{\mathbf{k}}^{d})_{B_{1g}}$ is peaked at $(\pi,0)$ and $(0,\pi)$.
Consequently,  $\Gamma _{\mathbf{k},B_{2g}}$ is strongly peaked at
$(\pi/2,\pi/2)$, while due to the mismatch of $\sin 2\theta _{\mathbf{k}}$ and
$(\gamma_{\mathbf{k}}^{c}+\gamma_{\mathbf{k}}^{d})_{B_{1g}}$,
$\Gamma_{\mathbf{k},B_{1g}}$ turns out to have two peaks located at
$(k_x,k_y)=(0.26\pi,0.74\pi)$ and $(0.74\pi,0.26\pi)$.
The symmetry mismatch between $\sin 2\theta _{\mathbf{k}}$ and
$(\gamma_{\mathbf{k}}^{c}+\gamma_{\mathbf{k}}^{d})_{B_{1g}}$
implies that the overall normal-state $B_{1g}$ Raman intensity should be {\em weaker}
than that of the $B_{2g}$ channel. The reported normal-state Raman intensities of
Ba(Fe$_{1-x}$Co$_x$)$_2$As$_2$ seem to be consistent with the prediction.
\cite{PhysRevB.80.180510}

Fig.~\ref{fig1}(f) shows the calculated normal-state Raman spectra in both
$B_{1g}$ and $B_{2g}$ channels. Apart from the feature that
$B_{1g}$ Raman intensity is much weaker than that of the $B_{2g}$ channel,
$B_{2g}$-channel Raman intensity shows a peak at $\omega\simeq 6.8 t_1$
while $B_{1g}$-channel Raman intensity shows a peak at $\omega\simeq 7.3 t_1$.
According to Eq.~(\ref{Raman response in normal state}), the frequencies
where the peaks appear are actually predictable.
When temperature $T\rightarrow 0$, the Raman peak for each channel can be
estimated to be at $\omega=E_+(k_x,k_y)-E_-(k_x,k_y)$ with
$(k_x,k_y)$ corresponding to the maximum of the weighting factor $\Gamma_{\bf k}$.
Therefore for $B_{2g}$-channel Raman intensity, the peak is predicted to be
at $\omega=E_+(\pi/2,\pi/2)-E_-(\pi/2,\pi/2)=6.8 t_1$,
while the $B_{1g}$-channel Raman peak is predicted to be at
$\omega=E_+(0.74\pi,0.26\pi)-E_-(0.74\pi,0.26\pi)=7.3 t_1$.
The above predicted normal-state Raman peaks may provide an
alternative route to the measurement of the band energy scale for
iron-pnictide superconductors.

In iron-pnictide superconductors, the energy scale of the nearest hopping
$t_1$ is estimated to be about $0.05 - 0.3$eV.
Thus the predicted normal-state Raman peaks could have energy $\omega\sim 0.35 - 2.1$eV.
The lower-bound energy (0.35 eV) should well be in the non-resonant regime,
while the upper-bound energy (2.1 eV) could be in the resonant regime and poses a
question mark for the above results to be valid.
Current reported normal-state Raman data on
BaFe$_{1-x}$Co$_x$)$_2$As$_2$ have only be measured up to 300cm$^{-1}$ however.
\cite{PhysRevB.80.180510}

\subsection{Superconducting state}  \label{sec32}

We next study the Raman spectra in the SC state. To do so, in a mean-field level
one can add a SC pairing Hamiltonian:
\begin{equation}
H_{\text{SC}}\ =\sum_{\mathbf{k}}\left( \Delta_{\mathbf{k}}^{\alpha}\alpha_{%
\mathbf{k}\uparrow}^{\dag}\alpha_{-\mathbf{k}\downarrow}^{\dag}+\Delta_{%
\mathbf{k}}^{\beta}\beta_{\mathbf{k}\uparrow}^{\dag}\beta _{-\mathbf{k}%
\downarrow}^{\dag}+\text{h.c.}\right)
\label{Hsc}
\end{equation}
to the diagonalized Hamiltonian $H_0$ in
(\ref{diagnalization Hamiltonian}).
In Eq.~(\ref{Hsc}), the pairing is considered between the long-lived
$\alpha_{\mathbf{k}\sigma}$ QPs and between $\beta_{\mathbf{k}\sigma}$ QPs only.
That is, interband pairing is neglected.
Since decoupled $\alpha$ and $\beta$ bands are originated from the
coupled $d_{xz}$ and $d_{yz}$-orbitals, the kind of SC pairing Hamiltonian (\ref{Hsc})
automatically includes both intra- and
inter-orbital pairings in the original fermion basis $\left( c_{%
\mathbf{k}\sigma},d_{\mathbf{k}\sigma}\right)$. \cite{PhysRevB.79.020501,
Nazario:144513}  The QP excitation energy is then given by
$\tilde{E}_{\mathbf{k}l}=(E^2_{l}+|\Delta_{\mathbf{k}}^l|^2)^{1/2}$ ($l=\alpha,\beta$)
for the two bands respectively. In this section, for convenience,
$E_-\rightarrow E_\alpha$ and $E_+\rightarrow E_\beta$.

\begin{figure}[ptb]
\begin{center}
\includegraphics[
width=9cm ]{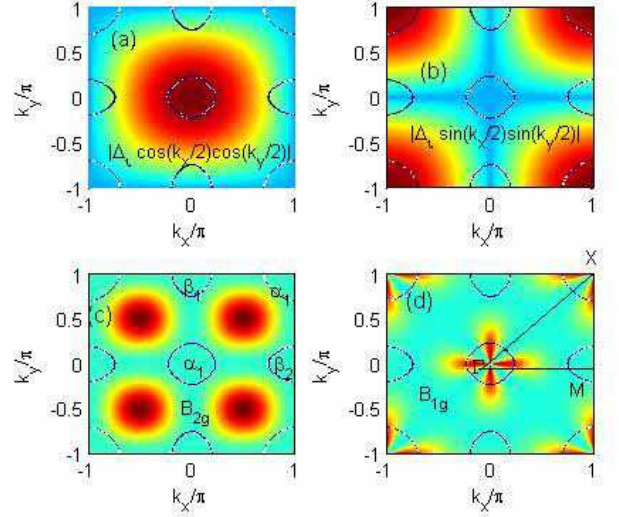}\vspace{-0.5cm}
\end{center}
\caption{(Color online) Panel (a)\&(b): Schematic plot of the $\alpha$- and $\beta$-band
Fermi surfaces and the $\mathbf{k}$-dependent amplitude
of (a) extended $s$-wave gap and (b) $d_{xy}$-wave gap. Panel (c)\&(d):
Momentum dependence of the effective Raman vertices
of (c) $(\gamma_{\bf k}^{\alpha\alpha})_{B_{2g}}$ and (d)
$(\gamma_{\bf k}^{\alpha\alpha})_{B_{1g}}$. Here we only show
$(\gamma_{\bf k}^{\alpha\alpha})_{B_{2g}}$ and $(\gamma_{\bf k}^{\alpha\alpha})_{B_{1g}}$ because
$(\gamma_{\bf k}^{\beta\beta})_{B_{2g}}=(\gamma_{\bf k}^{\alpha\alpha})_{B_{2g}}$ and
$(\gamma_{\bf k}^{\beta\beta})_{B_{1g}}\simeq (\gamma_{\bf k}^{\alpha\alpha})_{B_{1g}}$
(see text).}
\label{fig2}
\end{figure}

If the pairing originates from the same mechanism, most likely $\alpha_1$
and $\alpha_2$ bands will have the same pairing symmetry.
Similarly $\beta_1$ and $\beta_2$ bands will also likely have the same
pairing symmetry. The Raman spectra reported in
Ref.~[\onlinecite{PhysRevB.80.180510}] do not show a clear activation threshold
rather exhibit a finite intensity down to an arbitrarily small Raman shift.
It gives a strong evidence that the SC pairing
favors an anisotropic nodal gap rather than a full isotropic gap
such as the $s_\pm$ state or the $s_{++}$ state.
Furthermore, the scanning SQUID microscopy measurements seemed to
exclude the spin-triplet pairing states and suggested that the order parameter
has well-developed nodes [\onlinecite{hicks-2008}].
Within the {\em anisotropic} and {\em nodal} scenarios,
the possible candidates are the extended $s$-wave and  $d$-wave states.
\footnote{When the pairing symmetry in the unfolded BZ is extended $s$-wave with $\Delta_{\textbf{k}}=\Delta_0\cos\frac{k_x}{2}\cos\frac{k_y}{2}$ [see Fig.~\ref{fig2}(a)],
it will transform to be $\Delta_{\textbf{k}}=\Delta_0[\cos(k_x)+\cos(k_y)]$
in the folded BZ. Similarly, when the pairing symmetry in the unfolded BZ is
extended $d_{xy}$-wave with $\Delta_{\textbf{k}}=\Delta_0\sin\frac{k_x}{2}\sin\frac{k_y}{2}$
[see Fig.~\ref{fig2}(b)], it will transform to be
$\Delta_{\textbf{k}}=\Delta_0[\cos(k_x)-\cos(k_y)]$ in the folded BZ.}

After a lengthy derivation,
the irreducible Raman response function in the SC state is solved to be
\begin{eqnarray}
\chi_S(\mathbf{q}\rightarrow0,\tau)&=&-\sum_{\mathbf{k},ll^{\prime}=\alpha,\beta}
(\gamma_{\mathbf{k}}^{ll^{\prime}})^{2}[\mathcal{G}_{l}(\mathbf{k},\tau)\mathcal{G}%
_{l^{^{\prime}}}(\mathbf{k},-\tau)  \notag \\
& +&\mathcal{F}_{l}(\mathbf{k},\tau)\mathcal{F}_{l^{^{\prime}}}(\mathbf{k}%
,-\tau)],  \label{response
function}
\end{eqnarray}
where $\mathcal{G}_l$ and $\mathcal{F}_l$ are the usual normal and anomalous Green
functions for a superconductor associated with band $l$.
The intra- and interband vertex functions are solved to be
\begin{align}
\gamma_{\mathbf{k}}^{\alpha\alpha} & =\gamma_{\mathbf{k}}^{c}\cos^{2}\theta_{%
\mathbf{k}} +\gamma_{\mathbf{k}}^{d}\sin^{2}\theta_{\mathbf{k}},  \notag \\
\gamma_{\mathbf{k}}^{\beta\beta} & =\gamma_{\mathbf{k}}^{c}\sin^{2}\theta_{%
\mathbf{k}} +\gamma_{\mathbf{k}}^{d}\cos^{2}\theta_{\mathbf{k}},
\label{vertex functions} \\
\gamma_{\mathbf{k}}^{\alpha\beta} & =\gamma_{\mathbf{k}}^{\beta\alpha}=
\sin2\theta_{\mathbf{k}}(\gamma_{%
\mathbf{k}}^{c}+\gamma_{\mathbf{k}}^{d})/2,  \notag
\end{align}
where $\gamma_{\mathbf{k}}^{c}$ and $\gamma_{\mathbf{k}}^{d}$ were defined in (\ref{vertex})
for both $B_{1g}$ and $B_{2g}$ channels.
Frequency and channel-dependent Raman intensity is proportional to the imaginary part
of the effective density-density correlation function (\ref{response
function}) transformed to the Matsubara space.
As shown in Eq.~(\ref{response
function}), Raman spectra are contributed by both intraband and interband transitions.
Nevertheless, due to the little nesting effect occurring across different bands,
interband transitions are negligibly small for the Raman intensity.
%The strength of the interband vertices $\gamma_{\bf k}^{\alpha\beta}$ and $
%\gamma_{\bf k}^{\beta\alpha}$ is found to be much weaker than that of
%the intraband vertices, $\gamma_{\bf k}^{\alpha\alpha}$ and $\gamma_{\bf k}^{\beta\beta}$.
%In view of (\ref{vertex functions}), it is easily seen that the weakness
%of the interband vertices is due to the
%same that causes the weak Raman spectra in the normal state.
Thus one can safely ignore the interband Raman scattering
in the present case. Consequently, at $T\rightarrow 0$,
Raman intensity is proportional to
\begin{equation}\label{Raman intensity}
I(\omega) = \sum_{\mathbf{k},l=\alpha,\beta}(\gamma_{\bf k}^{ll})^2
\left({|\Delta_{\mathbf{k}}^l|\over 2\tilde{E}_{\mathbf{k}l}}\right)^2
\frac{\Gamma}{(\omega-2\tilde{E}_{\mathbf{k}l})^2+\Gamma^2},
\end{equation}
where $\Gamma$ is the broadening which is
set to be 0.08 in our calculations.

As is well-known, Raman scattering is a directional probe for SC
QP excitations. In the present two-band iron-pnictide superconductors,
the directional selectivity is dependent of two factors [see (\ref{Raman intensity})].
One is due to the
Raman vertex $\gamma_{\bf k}^{ll}$ and the other is due to
the symmetry of the pairing gap $\Delta_{\mathbf{k}}^l$.
The overall Raman will also depend on the detailed
locations and topology of the Fermi surfaces of the system.
It is worth noting that, as shown explicitly in (\ref{Raman intensity}),
Raman intensity is directly proportional to the gap maximum of the pairing gap.

\begin{figure}[ptb]
\begin{center}
\includegraphics[width=9cm ]{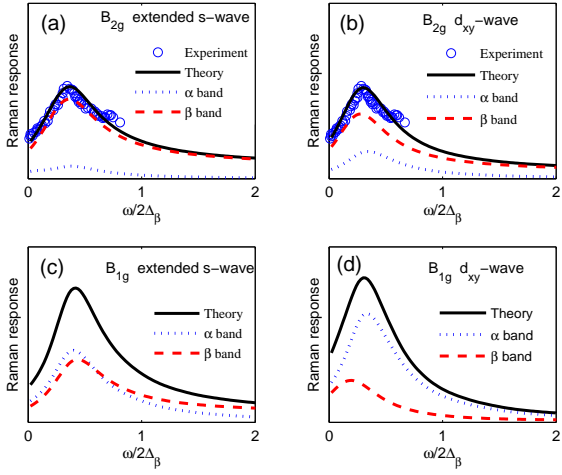}\vspace{-0.5cm}
\end{center}
\caption{Frame (a) and (b): Comparison of theoretical fitting and experimental data of the
$B_{2g}$ Raman spectra. The SC gaps tested are extended $s$-wave
$\Delta_{\bf k}=\Delta_{0}\cos k_{x}/2\cos k_{y}/2$ in (a)
and extended $d$-wave $\Delta_{\bf k}=\Delta_{0} \sin k_{x} \sin k_{y}$ in (b). Frame (c) and (d)
are the calculated $B_{1g}$ Raman shift using the same parameters
in frame (a) and (b).}
\label{fig3}
\end{figure}

It is important to first observe how the directional selection occurs
for the iron-pnictide superconductors. We consider
the unfolded BZ for the case of one Fe/cell.
The $\alpha$-band FSs of the 2-orbital model are hole Fermi pockets given by $
E_{-}(\mathbf{k}_{f}) = 0$ which are around the $\Gamma$ point and the
corner, $(\pm\pi,\pm\pi)$. The $\beta$-band FSs are electron Fermi
pockets given by $E_{+} (\mathbf{k}_{f}) = 0$ which are around the $M$
point. In the SC state. the coupling of the two orbitals results in complex Raman vertices
given in (\ref{vertex functions}).
Shown in Fig.~\ref{fig2} (c) and (d) are the momentum dependence of the Raman
vertices $(\gamma_{\bf k}^{\alpha\alpha})_{B_{2g}}$ and $(\gamma_{\bf k}^{\alpha\alpha})_{B_{1g}}$.
Note that for the current two-band model,
$(\gamma_{\bf k}^{\beta\beta})_{B_{2g}}=(\gamma_{\bf k}^{\alpha\alpha})_{B_{2g}}$ and
$(\gamma_{\bf k}^{\beta\beta})_{B_{1g}}\simeq (\gamma_{\bf k}^{\alpha\alpha})_{B_{1g}}$.
It is because
$(\gamma_{\bf k}^{c})_{B_{2g}}=(\gamma_{\bf k}^{d})_{B_{2g}}$ and
$(\gamma_{\bf k}^{c})_{B_{1g}} \simeq (\gamma_{\bf k}^{d})_{B_{1g}}$
[see Eqs.~(\ref{vertex}) and  (\ref{vertex functions})] for the current parameters.
As shown in Fig.~\ref{fig2} (c) and (d), $B_{2g}$ vertex peaks at $(\pm \pi/2, \pm\pi/2)$ in
the unfolded BZ and the peak is roughly of same distance to both
$\protect\alpha$- and $\protect\beta$-band FSs. As a matter of the fact,
it couples roughly equal to both $\protect\alpha$ and $\protect\beta$ bands.
%We therefore argue that the $B_{2g}$ channel will
%in principal probes both $\alpha$ and $\beta$ bands in the present two-band model.
In contrast, the $B_{1g}$ vertex has a $d_{x^2-y^2}$ symmetry and is centered
around the $\Gamma$ point and the corners, $(\pm\pi,\pm\pi)$.
Thus it couples predominantly to the $\alpha$ bands in the unfolded BZ.
Consequently $B_{1g}$ Raman scattering mainly excite the QP in the $\alpha$ bands and
can give more information about the pairing symmetry in the $\alpha$ bands.

We now test the possible extended $s$-wave and $d$-wave pairings.
Our approach is the following. We will try to fit the currently available $B_{2g}$
Raman intensity \cite{PhysRevB.80.180510} for each possible pairing symmetry.
The best fitting parameters of the gap magnitudes will be quoted. Using the
best fitting parameters for $B_{2g}$ Raman spectra, the predicted $B_{1g}$ Raman
intensity will be given. While $B_{1g}$ Raman intensity is also reported in
Ref.~[\onlinecite{PhysRevB.80.180510}], a strong phonon mode has appeared at $\omega=214$ cm$^{-1}$
responsible for Fe vibration, which makes the fitting unfeasible at the present time.

In our following calculations,
the only fitting parameters are $\Delta_{\alpha}$ and $\Delta_{\alpha}$ which are in units
of $t_1$. Both $\Delta_{\alpha}$ and $\Delta_{\beta}$ are adjusted to obtain
the best fitting for the experimental $B_{2g}$ spectra.
In particular, the $B_{2g}$ peak obtained through fitting
is identified with the ratio of $\omega/2\Delta_\beta$
which in turn is compared to the actual experimental
data of $\omega=69$cm$^{-1}$. One thus obtains the fitting value of $\Delta_\beta$.
With the knowledge of $\Delta_\beta$, one can further deduce the fitting values
of $\Delta_\alpha$ and $t_1$. The results for all possible pairing candidates
associated with extended $s$-wave and $d$-waves are
listed in Table~\ref{table1}.

\begin{figure}[ptb]
\begin{center}
\includegraphics[width=9cm ]{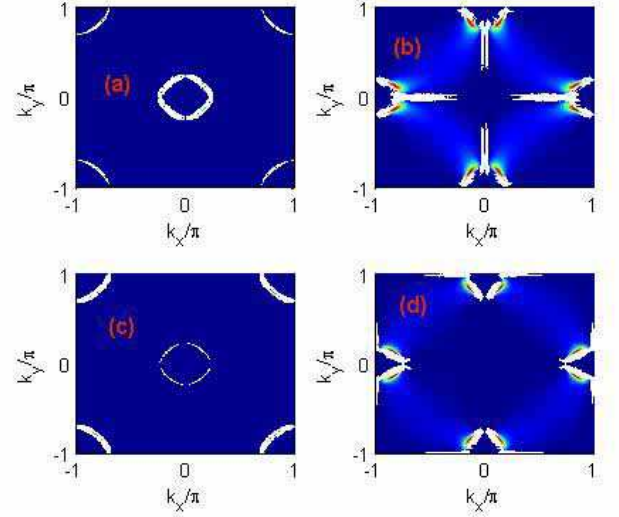}\vspace{-0.5cm}
\end{center}
\caption{Plot of the integrand, $(\gamma_{\bf k}^{ll})^2
({|\Delta_{\mathbf{k}}^l|/2\tilde{E}_{\mathbf{k}l}})^2$,
in Eq.~(\ref{Raman intensity}) in the first BZ.
Frame (a)\& (b) correspond respectively to $l=\alpha$ and $\beta$ with extended $s$-wave gap;
frame (c)\& (d) correspond respectively to $l=\alpha$ and $\beta$ with $d_{xy}$-wave gap.}
\label{fig4}
\end{figure}

We first consider the case of the extended $s$-wave pairing:
$\Delta_{\bf k}^l=\Delta_l\cos(k_x/2)\cos(k_y/2)$ ($l=\alpha,\beta$)
shown in Fig.~\ref{fig2}(a).
%In this case, due to the fact that gap amplitude near $\alpha_1$ FS is
%larger than that near the $\alpha_2$ FS,
%$\alpha_1$-band will contribute more to the Raman intensity than the $\alpha_2$-band.
%In contrast, $\beta_1$ and  $\beta_2$ bands will have the same contribution to the Raman intensity.
In a close observation of the $B_{2g}$ Raman spectra reported in
Ref.~\cite{PhysRevB.80.180510}, it is identified that only a {\em single}
peak develops at $\omega=69$ cm$^{-1}$. This is an important point in terms of
theoretical fitting.
In our fitting, the key is thus to ensure that both bands give a peak at the same frequency
($\omega=69$ cm$^{-1}$). {\em This one-peak scenario of fitting may seem unrealistic, but
indeed it is the only way to successfully describe the presently available data.}
It is found that for the best fitting [see Fig.~\ref{fig3}(a)],
$\Delta_\alpha/\Delta_\beta=0.35$.
The Raman peak occurs at $\omega/{2\Delta_\beta}=0.38$ which corresponds to
gap magnitudes $\Delta_\beta=91$ cm$^{-1}$ and $\Delta_\alpha=32$ cm$^{-1}$. Moreover
we have obtained $t_1=455$cm$^{-1}$.
%While a larger contribution due to the $\beta$-band because of
%a larger $\Delta_\beta$, it is in fact a smaller gap size near the $\beta$-band FS
%which determines the small-frequency Raman peak.

In view of Fig.~\ref{fig3}(a) for the best fitted curve, it is seen
that $\beta$-band contributes most to the overall $B_{2g}$ Raman intensity.
As mentioned before, $B_{2g}$ Raman vertex couples roughly equal
to the $\alpha$ and $\beta$ bands, thus the individual contribution to the Raman intensity
will depend crucially on the multiple effect of Raman vertex, symmetry and magnitude
of the gap function, as well as the Fermi surface topology. To see this multiple effect,
we have plotted in Fig.~\ref{fig4} the integrand, $(\gamma_{\bf k}^{ll})^2
({|\Delta_{\mathbf{k}}^l|/2\tilde{E}_{\mathbf{k}l}})^2$,
in Eq.~(\ref{Raman intensity}) in the first BZ.
It is shown that $\beta$-band has much stronger intensity near the $\beta$-band
FS which results in a
much stronger contribution to the overall Raman scattering [see Fig.~\ref{fig3}(a)].
Moreover, due to the nature of an extended $s$-wave gap which has nodes at the BZ edges,
Raman intensity is linear dependent at small frequencies.

The case of the $d_{xy}$-wave pairing in the unfolded BZ:
$\Delta_{\bf k}^l=\Delta_l\sin(k_x/2)\sin(k_y/2)$ ($l=\alpha,\beta$)
is studied next. As shown in Fig.~\ref{fig2}(b), the gap amplitude in
$\alpha_2$ FS is larger than that in $\alpha_1$ FS. Thus
the Raman intensity due to the $\alpha$-band will mainly contributed by the
$\alpha_2$-band FS. Due to the same gap amplitude near $\beta_1$ and $\beta_2$ FSs,
they will contribute equally to the Raman intensity.
In a similar approach, to obtain the same Raman peak for the two-band model,
we again use the two-gap approach. It is found that
$\Delta_\alpha/\Delta_\beta=0.3$ will give the best fitting for the experimental $B_{2g}$ data.
Moreover the $B_{2g}$ Raman peak corresponds to $\omega/{2\Delta_\beta}=0.3$
which in turn gives $\Delta_\beta=115$ cm$^{-1}$, $\Delta_\alpha=34$ cm$^{-1}$,
and $t_1=575$cm$^{-1}$.
While $\alpha$ band FSs are fully gapped, $\beta$-band FSs are gapped with a node.
Therefore the Raman intensity is powers-law dependent at low frequencies.
Moreover, as $\beta$-band contributes most to the Raman intensity because of larger gap size,
the low-energy $B_{2g}$ Raman intensity is actually linear.

The fitting to the Raman response in $B_{1g}$ channel is not feasible at the moment.
There occurs a strong phonon mode at 214 cm$^{-1}$ due to the
Fe vibration \cite{PhysRevB.80.180510}. The phonon mode is expected to be removed
by changing the crystal structure slightly.
We have calculated and predicted the $B_{1g}$ Raman responses for the extended $s$-wave
pairing symmetry shown in Fig.~\ref{fig3}(c) with the same parameters as those used in
Fig.~\ref{fig3}(a).
As shown in Fig.~\ref{fig2}(d), although $B_{1g}$ mode couples predominantly to the
$\alpha$ band in XM directions, the gap amplitude of
$\alpha$-band is smaller than that of the $\beta$-band. Consequently
it results QP excitation from the two bands with the same weight approximately.
The low-energy Raman response is predicted to be power-law dependent
due to the full gap in both FSs.

We have also calculated and predicted the $B_{1g}$ Raman responses
for the $d_{xy}$-wave pairing symmetry shown in Fig.~\ref{fig3}(d)
with the same parameters as those used in Fig.~\ref{fig3}(b).
As shown in Fig.~\ref{fig2}(b), $\alpha_1$ FS is near gap node while $\alpha_2$
FS is fully gapped. Thus $\alpha_1$-band will contribute more to the low-energy
Raman scattering than the $\alpha_2$-band. Since the gap amplitudes in  $\beta_1$ FS
is equal to that of $\beta_2$ FS, they have the same contribution to Raman scattering.
Although the gap amplitude of $\alpha$-band is smaller than that of the $\beta$-band,
$B_{1g}$ mode couples predominantly to the $\alpha$ band in XM directions however.
It results that $\alpha$ bands contribute more to the Raman scattering than
the $\beta$ band. The low-energy Raman response is predicted
to be power-law dependent with frequency due to the full gap in $\alpha_1$ band FS.

\begin{table}[ptb]
\caption{Summary of the fitting results for the three possible pairing symmetries on
the $B_{2g}$ Raman intensity.\cite{PhysRevB.80.180510}
Row 1 corresponds to the energy ratio of the  $B_{2g}$ peak (occurs at
69cm$^{-1}$) to $2\Delta_\beta$. Row 2 and 3 are the gap amplitudes
of $\Delta_\beta$ and $\Delta_\alpha$ in units of cm$^{-1}$ and
row 4 are their ratios. Row 5 are the deduced values of $t_1$
in units of cm$^{-1}$.}
\begin{ruledtabular}
\begin{tabular}{cccc}
Pairing  Symmetry & extended $s$-wave & $d_{xy}$-wave  & $d_{x^2-y^2}$-wave\\
  \hline
 ${\omega}/{2\Delta_\beta}$& 0.38 & 0.3 & 0.24 \\
 $\Delta_{\beta}$(cm$^{-1}$) & 91 & 115 & 143\\
 $\Delta_{\alpha}$(cm$^{-1}$) & 32 & 34 & 14\\
 $\Delta_{\alpha}/\Delta_\beta$& 0.3 & 0.3 & 0.1  \\
$t_1$(cm$^{-1}$)& 455 & 575 & 7150  \\
\end{tabular}
\end{ruledtabular}
\label{table1}
\end{table}

To complete the studies, we have also studied the case of $d_{x^2-y^2}$-wave pairing symmetry.
It is found that the best fitting of the $B_{2g}$ Raman intensity is given by
the ratio $\Delta_{\alpha}/\Delta_{\beta}=0.1$ which in turn gives $t_1=7150$cm$^{-1}$.
Based on the given unrealistically large $t_1$,
we conclude that $d_{x^2-y^2}$-wave pairing is ruled out in terms of the current available
Raman scattering data. Table~\ref{table1} is a summary of the
the fitting results for the  $B_{2g}$ Raman intensity.

\section{Summary} \label{Summary}
In summary, we have studied the Raman response of iron-pnictide
superconductor in both normal and SC states based on a two-band model.
Predictions are given for the normal-state Raman intensities.
A more quantitative fitting to the currently available Raman spectra is made
to which useful fitting parameters are quoted in terms of the
gap amplitudes on both bands.

%The calculation of Raman scattering explain that the Raman intensity in normal state is very weak due to the
%mismatch of the Raman vertex and the interband transition.  We use various possible SC pairing to fit Raman response %in $B_{2g}$ channel and find the extended $s$-wave with $\cos(k_x)-\cos(k_x)$ pairing symmetry and $d_{xy}$-wave quite %well.
%The small Raman shift is further explained by the interplay between multi-Fermi surface and isotropic SC pairing, as %well as vertex correlation. We further argue that present Raman date support more the pairing symmetry is $d_{xy}$ %wave and the extended wave cannot be excluded. To understand the SC pairing  of Fe-pnictides
%superconductor, it is desired a more Raman data and a more comprehensive theory including many orbital.

\begin{acknowledgments}
This work was supported by National Science Council of Taiwan (Grant No.
99-2112-M-003-006), Hebei Provincial Natural Science Foundation of China (Grant
No. A2010001116), and the National Natural Science
Foundation of China (Grant No. 10974169). We also acknowledge the
support from the National Center for Theoretical Sciences, Taiwan.
\end{acknowledgments}

%\bibliography{Ref}
%Merlin.mbs v4.21 2009-07-09.
%

\end{document}